\documentclass[
    ,final            
  ]
  {aipproc}

\layoutstyle{6x9}

\begin{document}

\title{New supersymmetric quartet of nuclei: $^{192,193}$Os-$^{193,194}$Ir}


\author{R. Bijker}{
  address={ICN-UNAM, AP 70-543, 04510 M\'exico DF, M\'exico}
}

\author{J. Barea}{
  address={Center for Theoretical Physics, Sloane Physics Laboratory, 
Yale University, P.O. Box 208210, New Haven, Connecticut 06520-8120, U.S.A.}
}

\author{A. Frank}{
  address={ICN-UNAM, AP 70-543, 04510 M\'exico DF, M\'exico}
}

\author{G. Graw}{
  address={Sektion Physik, Ludwig-Maximilians-Universit\" at M\" unchen, D-85748 Garching, Germany}
}

\author{R. Hertenberger}{
  address={Fakult\"at f\"ur Physik, Ludwig-Maximilians-Universit\" at M\" unchen, 
D-85748 Garching, Germany}
}

\author{J. Jolie}{
  address={Institut f\" ur Kernphysik, Universit\" at zu K\" oln, D-50937 K\" oln, Germany}
}

\author{H.-F. Wirth}{
  address={Sektion Physik, Ludwig-Maximilians-Universit\" at M\" unchen, D-85748 Garching, Germany}
}

\begin{abstract}
We present evidence for the existence of a new supersymmetric quartet of nuclei in the 
$A \sim 190$ mass region. The analysis is based on new experimental information on the 
odd-odd nucleus $^{194}$Ir from transfer and neutron capture reactions. The new data 
allow the identification of a new supersymmetric quartet, consisting of the $^{192,193}$Os 
and $^{193,194}$Ir nuclei. We make explicit predictions for $^{193}$Os, and suggest that 
its spectroscopic properties be measured in dedicated experiments. 
Finally, we study correlations between different transfer reactions.
\end{abstract}

\maketitle

\section{Introduction}

The $A \sim 190$ mass region is a particularly complex one, displaying transitional 
behavior such as prolate-oblate deformed shapes, $\gamma$-unstability, triaxial deformation 
and/or coexistence of different configurations which present a daunting challenge to 
nuclear structure models. Despite this complexity, the $A \sim 190$ mass region has 
been a rich source of empirical evidence for the existence of dynamical symmetries in 
nuclei both for even-even, odd-proton, odd-neutron and odd-odd nuclei, as well as 
supersymmetric pairs \cite{FI,baha} and quartets of nuclei \cite{quartet,metz}. 

In this contribution, we present evidence for the existence of a new supersymmetric 
quartet in the $A \sim 190$ mass region, consisting of the $^{192,193}$Os and $^{193,194}$Ir 
nuclei, and study correlations between different one- and two-nucleon transfer reactions. 

\section{Nuclear supersymmetry}

Dynamical supersymmetries (SUSY) were introduced in nuclear physics in 
the context of the Interacting Boson Model (IBM) and its extensions \cite{FI}.  
The IBM describes collective excitations in even-even nuclei in 
terms of a system of interacting monopole ($s^{\dagger}$) and quadrupole 
($d^{\dagger}$) bosons \cite{IBM}. The bosons are associated with the number of 
correlated proton and neutron pairs, and hence the number of bosons $N$ is 
half the number of valence nucleons.  
For odd-mass nuclei the IBM was extended to include single-particle
degrees of freedom \cite{olaf}. The ensuing Interacting Boson-Fermion Model 
(IBFM) has as its building blocks $N$ bosons with $l=0,2$ and $M=1$ fermion 
($a_j^{\dagger}$) with $j=j_1,j_2,\dots$ \cite{IBFM}. The IBM and IBFM can 
be unified into a supersymmetry (SUSY) $U(6/\Omega) \supset U(6) \otimes U(\Omega)$
where $\Omega=\sum_j (2j+1)$ is the dimension of the fermion space \cite{FI} . 
In this framework, even-even and odd-even nuclei form the members of a 
supermultiplet which is characterized by ${\cal N}=N+M$, 
i.e., the total number of bosons and fermions.  
Supersymmetry distinguishes itself from other symmetries in that it includes, 
in addition to transformations among fermions and among bosons, also 
transformations that change a boson into a fermion and {\em vice versa}.

The concept of nuclear SUSY was extended in 1985 to include the neutron-proton 
degree of freedom \cite{quartet}. In this case, a supermultiplet consists of an 
even-even, an odd-proton, an odd-neutron and an odd-odd nucleus. The 
the best experimental evidence of a supersymmetric quartet is provided by the 
$^{194,195}$Pt and $^{195,196}$Au nuclei as an example of the 
$U_{\nu}(6/12) \otimes U_{\pi}(6/4)$ supersymmetry \cite{metz,groeger,wirth,barea1,barea2},   
in which the odd neutron is allowed to occupy the $3p_{1/2}$, $3p_{3/2}$ and $2f_{5/2}$ 
orbits of the 82-126 shell, and the odd proton the $2d_{3/2}$ orbit of the 50-82 shell.  
This supermultiplet is characterized by ${\cal N}_{\nu}=5$ and ${\cal N}_{\pi}=2$. 
The excitation spectra of the nuclei belonging to the supersymmetric 
quartet are described simultaneously by the energy formula
\begin{eqnarray}
E &=& A \left[ N_1(N_1+5)+N_2(N_2+3)+N_1(N_1+1) \right] 
\nonumber\\
&& + B  \left[ \Sigma_1(\Sigma_1+4)+\Sigma_2(\Sigma_2+2)+\Sigma_3^2 \right] 
+ B' \left[ \sigma_1(\sigma_1+4)+\sigma_2(\sigma_2+2)+\sigma_3^2 \right] 
\nonumber\\
&& + C \left[ \tau_1(\tau_1+3)+\tau_2(\tau_2+1) \right] + D \, L(L+1) + E \, J(J+1) ~.
\label{npsusy}
\end{eqnarray}  
The coefficients $A$, $B$, $B'$, $C$, $D$, and $E$ are determined in a simultaneous 
fit of the excitation energies of the four nuclei that make up the quartet. 

Recently, the structure of the odd-odd nucleus $^{194}$Ir was investigated by a series 
of transfer and neutron capture reactions \cite{balodis}. In particular, the new data 
from the polarized $(\vec{d},\alpha)$ transfer reaction provided crucial new information 
about and insight into the structure of the spectrum of $^{194}$Ir which led 
to significant changes in the assignment of levels as compared to previous work
\cite{joliegarrett}. 

\begin{figure}
\includegraphics[width=80mm]{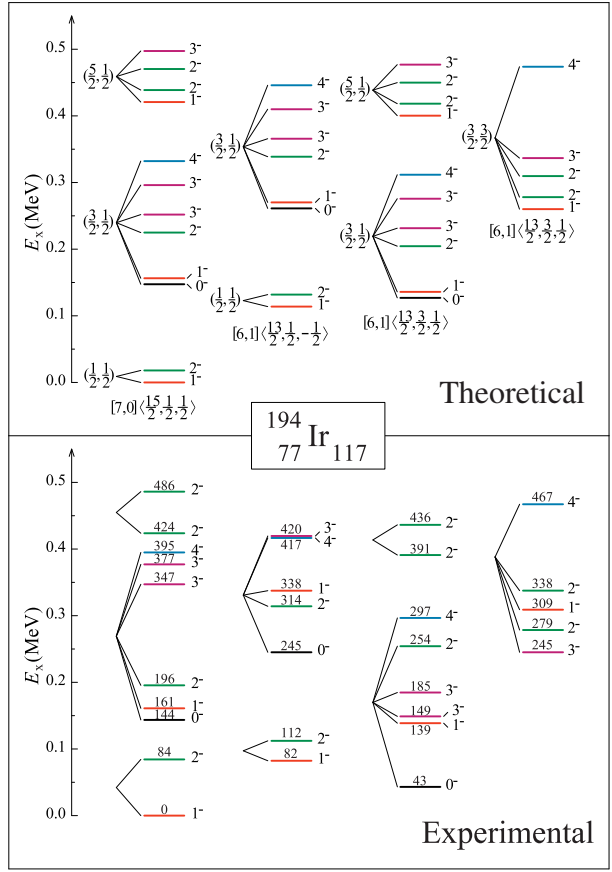} 
\caption{Comparison between the theoretical and experimental spectrum of $^{194}$Ir.}
\label{ir194} 
\end{figure}

The odd-odd nucleus $^{194}$Ir differs from $^{196}$Au by two protons, the 
number of neutrons being the same. The latter is crucial, since the dominant 
interaction between the odd neutron and the core nucleus is of quadrupole type, 
which arises from a more general interaction in the IBFM for very special values 
of the occupation probabilities of the $3p_{1/2}$, $3p_{3/2}$ and $2f_{5/2}$ 
orbits, {\em i.e.} to the location of the Fermi surface for the neutron orbits 
\cite{bijker}. This situation is satisfied to a good approximation by the 
$^{195}$Pt and $^{196}$Au nuclei, and thus also for $^{193}$Os and $^{194}$Ir. 
For this reason, it is reasonable to expect the odd-odd nucleus $^{194}$Ir 
to provide another example of the $U(6/12)_{\nu} \otimes U(6/4)_{\pi}$  
supersymmetry. Fig.~\ref{ir194} shows the negative parity levels of $^{194}$Ir 
in comparison with the theoretical spectrum in which it is assumed that these 
levels originate from the $\nu 3p_{1/2}$, $\nu 3p_{3/2}$, 
$\nu 2f_{5/2} \otimes \pi 2d_{3/2}$ configuration. 
The theoretical energy spectrum is calculated using the energy formula of 
Eq.~(\ref{npsusy}) with $A = 26.3$, $B = 8.7$, $B' = -33.6$, $C = 35.1$, 
$D = 6.3$, and $E = 4.5$ (all in keV). Given the complex nature of the spectrum 
of heavy odd-odd nuclei, the agreement is remarkable. There is an 
almost one-to-one correlation between the experimental and theoretical level 
schemes \cite{balodis}. 

\begin{figure}
\includegraphics[width=80mm]{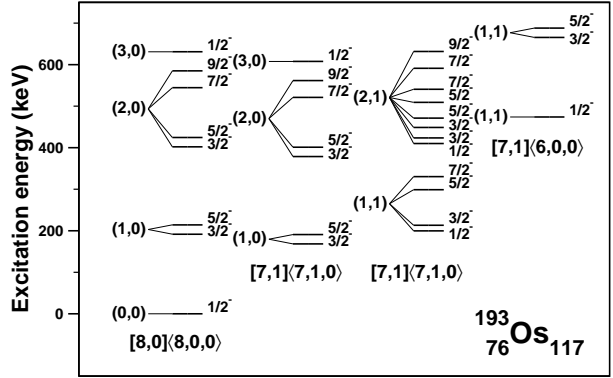} 
\caption{Prediction of the spectrum of $^{193}$Os for the 
$U_{\nu}(6/12) \otimes U_{\pi}(6/4)$ supersymmetry.}
\label{os193} 
\end{figure}

The successful description of the odd-odd nucleus $^{194}$Ir opens the possibility 
of identifying a second quartet of nuclei in the $A \sim 190$ mass region with 
$U(6/12)_{\nu} \otimes U(6/4)_{\pi}$ supersymmetry. The new quartet consists of the 
nuclei $^{192,193}$Os and $^{193,194}$Ir and is characterized by ${\cal N}_{\nu}=5$ 
and ${\cal N}_{\pi}=3$. Whereas the $^{192}$Os and $^{193,194}$Ir nuclei are well-known 
experimentally, the available data for $^{193}$Os is rather scarce. In Fig.~\ref{os193} 
we show the predicted spectrum for $^{193}$Os obtained from Eq.~(\ref{npsusy}) using 
the same parameter set as for $^{194}$Ir. We note, that the ground state of $^{193}$Os 
has spin and parity $J^P=\frac{3}{2}^{-}$, which implies that the second band with 
labels $[7,1]$, $\langle 7,1,0 \rangle$ is the ground state band, rather than 
$[8,0]$, $\langle 8,0,0 \rangle$. The relative ordering 
of these bands is determined by the coefficients $A$ and $B+B'$. At present, we are 
carrying out a simultaneous fit of the excitation energies of all four nuclei that 
make up the quartet to see whether it is possible to reproduce the relative ordering 
in $^{193}$Os without affecting the successful description of $^{194}$Ir \cite{osir}

\section{Correlations}

The nuclei belonging to a supersymmetric quartet are described by a single Hamiltonian, 
and hence the wave functions, transition and transfer rates are strongly 
correlated. As an example of these correlations, we consider here the case 
of one-neutron transfer reactions between the Pt and Os nuclei.    

In a study of the $^{194}$Pt $\rightarrow$ $^{195}$Pt stripping reaction 
it was found \cite{bi} that one-neutron $j=3/2$, $5/2$ transfer 
reactions can be described in the $U(6/12)_{\nu} \otimes U(6/4)_{\pi}$ 
supersymmetry scheme by the operator 
\begin{eqnarray}
P_{\nu}^{(j) \, \dagger} &=& \alpha_j \frac{1}{\sqrt{2}} \left[ 
\left( \tilde{s}_{\nu} \times a^{\dagger}_{\nu,j} \right)^{(j)} - 
\left( \tilde{d}_{\nu} \times a^{\dagger}_{\nu,1/2} \right)^{(j)} \right] ~. 
\end{eqnarray}
It is convenient to take ratios of intensities, since they do not depend on the 
value of the coefficient $\alpha_j$ and hence provide a stringent test of the wave 
functions. For the stripping reaction $^{194}$Pt $\rightarrow$ $^{195}$Pt 
(ee $\rightarrow$ on) the ratio of intensities for the excitation of the $(1,0)$, $L=2$ 
doublet with $J=3/2$, $5/2$ belonging to the first excited band with $[N+1,1]$, 
$(N+1,1,0)$ relative to that of the ground state band $[N+2]$, $(N+2,0,0)$ is given 
by \cite{bi} 
\begin{eqnarray}
R_j({\rm ee \rightarrow on}) = \frac{(N+1)(N+3)(N+6)}{2(N+4)} ~,
\label{ratio}
\end{eqnarray}
which gives $R_j=29.3$ for $^{194}$Pt $\rightarrow$ $^{195}$Pt ($N=5$), 
in comparison with the experimental value of 19.0 for $j=5/2$, and 
$R_j=37.8$ for $^{192}$Os $\rightarrow$ $^{193}$Os ($N=6$).   
The equivalent ratio for the inverse pick-up reaction is given by
\begin{eqnarray}
R_j({\rm on \rightarrow ee}) = R_j({\rm ee \rightarrow on}) 
\frac{N_{\pi}+1}{(N+1)(N_{\nu}+1)} ~. 
\label{corr}
\end{eqnarray}
which gives $R_j=1.96$ for $^{195}$Pt $\rightarrow$ $^{194}$Pt ($N_{\pi}=1$ and 
$N_{\nu}=4$) and $R_j=3.24$ for $^{193}$Os $\rightarrow$ $^{192}$Os ($N_{\pi}=2$ 
and $N_{\nu}=4$). This means that the mixed symmetry $L=2$ state is predicted 
to be excited more strongly than the first excited $L=2$ state.  
This correlation between pick-up and stripping reactions can be derived in a general 
way only using the symmetry relations that exist between the wave functions of the 
even-even and odd-neutron nuclei of the supersymmetric quartet. It is important 
to point out, that Eqs.~(\ref{ratio} and (\ref{corr}) are parameter-independent 
predictions which are a direct consequence of nuclear SUSY and which can be 
tested experimentally.  

\section{Summary and conclusions}

In conclusion, we have presented evidence for the existence of a second quartet 
of nuclei in the $A\sim 190$ region with $U_{\nu}(6/12)\otimes U_{\pi}(6/4)$ 
supersymmetry, consisting of the $^{192,193}$Os and $^{193,194}$Ir nuclei. The 
analysis is based on new experimental information on $^{194}$Ir. In particular,  
the $(\vec{d},\alpha)$ reaction is important to establish the spin and parity 
assignments of the energy levels, and to provide insight into the structure of 
the spectrum of $^{194}$Ir. Given the complexity of the $A \sim 190$ mass region, 
the simple yet detailed description of $^{194}$Ir in a supersymmetry scheme is 
truly remarkable.  

Nuclear supersymmetry establishes precise links among the spectroscopic properties 
of different nuclei. This relation has been used to predict the energies of 
$^{193}$Os. Since the wave functions of the members of a supermultiplet are connected 
by symmetry, there exists a high degree of correlation between different one- and 
two-nucleon transfer reactions not only between nuclei belonging to the same quartet, 
but also for nuclei from different multiplets \cite{barea1,barea2}.  
As an example, we studied the correlations between one-neutron transfer reactions 
for the Pt and Os isotopes, and predicted that the $L=2$ mixed symmetry states 
in the even-even nucleus are populated much stronger than the first excited $L=2$ state. 
 
In order to establish the existence of a second supersymmetric quartet of nuclei 
in the $A \sim 190$ mass region, it is crucial that the nucleus $^{193}$Os be studied 
in more detail experimentally. The predictions for correlations between one-neutron 
transfer reactions in Pt and Os can be tested experimentally by combining for example 
$(\vec{d},p)$ stripping and $(p,d)$ pick-up reactions.  

\begin{theacknowledgments}
This work was supported in part by PAPIIT-UNAM (grant IN113808), 
and in part by the Deutsche Forschungsgemeinschaft (grants JO391/2-3 and GR894/2-3). 
\end{theacknowledgments}


\begin{thebibliography}{99}

\bibitem{FI} 
F. Iachello, 
{\em Phys. Rev. Lett.} {\bf 44}, 772 (1980). 

\bibitem{baha}
A.B. Balantekin, I. Bars, R. Bijker and F. Iachello, 
{\em Phys. Rev.} C {\bf 27}, 1761 (1983).

\bibitem{quartet}
P. Van Isacker, J. Jolie, K. Heyde and A. Frank, 
{\em Phys. Rev. Lett.} {\bf 54}, 653 (1985).

\bibitem{metz}
A. Metz {\em et al.}, 
{\em Phys. Rev. Lett.} {\bf 83}, 1542 (1999).

\bibitem{IBM}
F. Iachello and A. Arima, 
{\em The interacting boson model}  
(Cambridge U. Press, 1987). 

\bibitem{olaf}
F. Iachello and O. Scholten,
{\em Phys. Rev. Lett.} {\bf 43}, 679 (1979).
 
\bibitem{IBFM}
F. Iachello and P. Van Isacker, 
{\em The interacting boson-fermion model}  
(Cambridge U. Press, 1991). 

\bibitem{groeger}
J. Gr\"oger {\em et al.}, 
{\em Phys. Rev.} C {\bf 62}, 064304 (2000).

\bibitem{wirth}
H.-F. Wirth {\em et al.}, 
{\em Phys. Rev.} C {\bf 70}, 014610 (2004).

\bibitem{barea1}
J. Barea, R. Bijker and A. Frank,  
{\em J. Phys. A: Math. Gen.} {\bf 37}, 10251 (2004). 

\bibitem{barea2}
J. Barea, R. Bijker and A. Frank, 
{\em Phys. Rev. Lett.} {\bf 94}, 152501 (2005). 

\bibitem{balodis}
M. Balodis {\it et al.}, 
{\em Phys. Rev.} C {\bf 77}, 064602 (2008).

\bibitem{joliegarrett} 
J. Jolie and P. Garrett, 
{\em Nucl. Phys.} A {\bf 596}, 234 (1996).

\bibitem{bijker}
R. Bijker and O. Scholten, 
{\em Phys. Rev.} C {\bf 32}, 591 (1985).

\bibitem{bi}
R. Bijker and F. Iachello, 
{\em Ann. Phys. (N.Y.)} {\bf 161}, 360 (1985).

\bibitem{osir}
J. Barea {\em et al.}, work in progress.

\end{thebibliography}
\end{document}